 \documentclass[final,5p,times,twocolumn]{elsarticle}

\usepackage{amssymb}

\journal{BBA - General Subjects}

\begin{document}

\begin{frontmatter}



\title{Broadband Dielectric Spectroscopy on Human Blood}


\author[label1]{M. Wolf}

\author[label1]{R. Gulich}

\author[label1]{P. Lunkenheimer\corref{cor1}}
\ead{peter.lunkenheimer@physik.uni-augsburg.de}
\cortext[cor1]{Corresponding author. Tel.: + 49 821 5983603  FAX +
49 821 5983649}

\author[label1]{A. Loidl}

\address[label1]{Experimental Physics V, Center
for Electronic Correlations and Magnetism, University of Augsburg,
86135 Augsburg, Germany}

\begin{abstract}

\bigskip\noindent\textbf{Background}

Dielectric spectra of human blood reveal a rich variety of dynamic
processes. Achieving a better characterization and understanding of
these processes not only is of academic interest but also of high
relevance for medical applications as, e.g., the determination of
absorption rates of electromagnetic radiation by the human body.

\bigskip\noindent\textbf{Methods}

The dielectric properties of human blood are studied using broadband
dielectric spectroscopy, systematically investigating the dependence
on temperature and hematocrit value. By covering a frequency range
from 1~Hz to 40~GHz, information on all the typical dispersion
regions of biological matter is obtained.

\bigskip\noindent\textbf{Results and conclusions}

We find no evidence for a low-frequency relaxation
("$\alpha$-relaxation") caused, e.g., by counterion diffusion
effects as reported for some types of biological matter. The
analysis of a strong Maxwell-Wagner relaxation arising from the
polarization of the cell membranes in the 1-100~MHz region
("$\beta$-relaxation") allows for the test of model predictions and
the determination of various intrinsic cell properties. In the
microwave region beyond 1~GHz, the reorientational motion of water
molecules in the blood plasma leads to another relaxation feature
("$\gamma$-relaxation"). Between $\beta$- and $\gamma$-relaxation,
significant dispersion is observed, which, however, can be explained
by a superposition of these relaxation processes and is not due to
an additional "$\delta$-relaxation" often found in biological
matter.

\bigskip\noindent\textbf{General significance}

Our measurements provide dielectric data on human blood of so far
unsurpassed precision for a broad parameter range. All data are
provided in electronic form to serve as basis for the calculation of
the absorption rate of electromagnetic radiation and other medical
purposes. Moreover, by investigating an exceptionally broad
frequency range, valuable new information on the dynamic processes
in blood is obtained.

\end{abstract}


\end{frontmatter}

\section{Introduction}
\label{intro} Blood is a highly functional body fluid, it delivers
oxygen to the vital parts, it transports nutrients, vitamins, and
metabolites and it also is a fundamental part of the immune system.
Therefore the precise knowledge of its constituents, its physical,
biological, and chemical properties and its dynamics is of great
importance. Especially its dielectric parameters are of relevance
for various medical applications \cite{Grant1978}, like cell
separation (e.g., cancer cells from normal blood cells
\cite{Becker1995}), checking the deterioration of preserved blood
\cite{Hayashi2010a}, and dielectric coagulometry \cite{Hayashi2010}.
In addition, the precise knowledge of the dielectric properties of
blood is prerequisite for fixing limiting values for electromagnetic
pollution (via the conductivity in the specific absorption rate
(SAR))\cite{Ahlbom1998, IEEE2002, IEEE2005, Loidl2008}.

Early measurements of the electrical properties of blood contributed
significantly to unravel the constitution of red blood cells (RBC).
For example, the results by H\"{o}ber \cite{Hoeber1913} provided the
first indications of a dispersion (i.e. frequency dependence),
caused by the membrane of RBCs, in the radio frequency (RF) spectrum
of the dielectric properties of blood. This relaxation process is
nowadays identified as being of Maxwell-Wagner type
\cite{Maxwell1873, Wagner1914} and termed $\beta$-relaxation in
biophysical literature \cite{Schwan1957a,Schwan1983}.

Various early works \cite{Fricke1923, Fricke1924, Fricke1925a,
Fricke1925b, Cole1928, Daenzer1934b, Bruggemann1935, Rajewsky1948}
were followed by measurements at very high frequencies
\cite{Schwan1953, England1950, Cook1951, Cook1952}. Some of them
revealed an additional dispersion with a relaxation rate near
18~GHz, which can be assigned to the reorientation of water
molecules and which is named $\gamma$-dispersion. Furthermore, a
third relaxation, termed $\alpha$-relaxation and located in the
low-frequency regime, $\nu<100$~kHz was detected in some biological
materials \cite{Schwan1954, Schwan1957}. However, interestingly, an
$\alpha$-relaxation seems to be absent in whole blood
\cite{Bothwell1956} and only is found in hemolyzed blood cells
\cite{Schwan1957}. This was speculated to be due to a higher ion
permeability of the membranes in the latter case, shifting the
relaxation spectrum into the experimental frequency window
\cite{Schwan1957}. The origin of the $\alpha$-relaxation is a matter
of controversy; most commonly it is assumed to arise from counterion
diffusion effects \cite{Schwan1983, Schwan1962}. Finally, a
dispersion with low dipolar strength located in the frequency regime
between the $\beta$- and $\gamma$-dispersion was identified by
Schwan \cite{Schwan1957a}. The origin of the $\delta$-dispersion and
the possible role of bound water in its generation is
controversially discussed \cite{Schwan1965, Grant1965, Grant1968,
Pennock1969, Nandi2000, Feldman2003, Knocks2001, Thomas2008,
Sasisanker2008}.

Taking together all these results, it is clear that there are three
main dispersion regions in the dielectric frequency spectrum of
blood between some Hz and 50~GHz, termed $\beta$, $\gamma$, and
$\delta$ \cite{Schwan1957a,Schwan1983}. This nomenclature should not
be confused with that used in the investigation of glassy matter
like supercooled liquids or polymers. Within the glass-physics
community the terms $\alpha$-, $\beta$-relaxation, etc. are commonly
applied to completely different phenomena than those considered
above (see, e.g., refs \citenum{Lunkenheimer2000} and
\citenum{Kremer2002}). In the present work we follow the biophysical
nomenclature.

A lot of additional research has been done until the early 1980s
\cite{Grant1978, Pauly1966, Krupa1972,  Schanne1978, Pethig1979} and
a detailed review was given in 1983 by Schwan \cite{Schwan1983}.
Later on, in the course of the upcoming debates about
electromagnetic pollution, dielectric properties of body tissues and
fluids received renewed interest as they determine the SAR, a
measure for the absorption of electromagnetic fields by biological
tissue \cite{Loidl2008, Gabriel1996a, Gabriel1996b, Foster1989,
Foster1995}. But also various other important medical questions can
be addressed by dielectric spectroscopy \cite{Grant1978, Becker1995,
Hayashi2010a, Hayashi2010, Markx1999}. In the last two decades, a
number of papers on dielectric spectroscopy on blood and erythrocyte
solutions were published \cite{Feldman2003, Lisin1996, Zhao1993,
Bordi1997, Jaspard2003, Beving1994a, Bao1994,  Chelidze2002,
Hayashi2008}, most of them treating special aspects only.

On the theoretical side, a number of models for the description of
cell suspensions have been proposed. Most models focus on the
$\beta$-relaxation \cite{Schwan1957a, Fricke1923, Fricke1924,
Fricke1925a, Fricke1925b, Bruggemann1935, Pauly1959,
Hanai1960a,Looyenga1965, Asami2002, Hanai1961, Hanai1968,
Katsumoto2008}, including the often employed Pauly-Schwan model
\cite{Schwan1957a, Pauly1959}, discussed below. Some of them also
account for the non-spherical shape of cells \cite{Fricke1923,
Fricke1924, Fricke1925a, Fricke1925b, Asami2002}. It seems clear
that diluted solutions and whole blood with a hematocrit value of
86\% have to be treated differently. The Bruggeman-Hanai model
\cite{Bruggemann1935, Hanai1960a} was specially developed for highly
concentrated suspensions. A recent summary of various models can be
found in ref. \cite{Asami2002}.

Concluding this introduction, it has to be stressed that, after more
than one century of research, many aspects of the dielectric
properties of blood (e.g., the presence and origins of $\alpha$- and
$\delta$-dispersion) are still unclear. It should be noted that, in
addition to the three main dispersion effects, from a theoretical
viewpoint a number of further relaxation features may show up in
blood. For example, it is well known that RBC's are far from being
of spherical shape and in principle for shelled ellipsoidal
particles up to six relaxations can be expected \cite{Asami2002}.
Furthermore, the hemoglobin molecules within the RBC's should show
all the typical complex dynamics as found in other proteins. Based
on the available literature data (e.g., \cite{Schwan1983}), it seems
that most, if not all, of these additional processes do not or only
weakly contribute to the experimentally observed spectra. However,
one should carefully check for possible deviations from the simple
three-relaxation scenario mentioned above, which may well be
ascribed to these additional processes.

Maybe the best and most cited broadband spectra of blood covering
several dispersion regions are those by Gabriel \textit{et al}
\cite{Gabriel1996b}, taken at 37 C, which are commonly used for SAR
calculations and for medical purposes. However, even these data are
hampered by considerable scatter and they are composed from data
collected by different groups on different samples. Clearly,
high-quality spectra covering a broad frequency range measured on
identical samples are missing. A systematic investigation of the
hematocrit and temperature dependence is essential to achieve a
better understanding of the different dispersion contributions of
blood. The present work provides the dielectric constant
$\varepsilon'$, the loss $\varepsilon''$, and the conductivity
$\sigma'$ of human blood in a broad frequency range (1~Hz to
40~GHz), by using a combination of different techniques of
dielectric spectroscopy applied to identical samples. In addition,
the temperature (280~K~-~330~K) and hematocrit value (0~-~86\%)
dependence is thoroughly investigated.

\section{Models and Data Analysis}
\label{Models}

Dielectric spectroscopy is sensitive to dynamical processes that
involve the reorientation of dipolar entities or displacement of
charged entities, which can cause a dispersive behavior of the
dielectric constant and loss. However, also non-intrinsic
Maxwell-Wagner effects caused by interfacial polarization in
heterogeneous samples can lead to considerable dispersion
\cite{Maxwell1873, Wagner1914, Lunkenheimer2002}. As mentioned
above, biological matter shows various dispersions in the frequency
regime 1~Hz to 40~GHz, which have different microscopic and
mesoscopic origins and therefore have to be described differently.

\subsection{Intrinsic Relaxations}
\label{Intrinsic}

Intrinsic processes like, e.g., the cooperative reorientation of
dipolar molecules are often described by the Debye formula
\cite{Debye1929}:

\begin{equation}
\label{Db}
\varepsilon^{*}(\nu)=\varepsilon_{\infty}+\frac{\Delta\varepsilon}{1+\mathrm{i}2\mathrm{\pi}\nu\tau}\:
\end{equation}

\noindent The relaxation strength is given by
$\Delta\varepsilon=\varepsilon_{\mathrm{s}}-\varepsilon_{\mathrm{\infty}}$.
$\varepsilon_{\mathrm{s}}$ and $\varepsilon_{\infty}$ are the
limiting values of the real part of the dielectric constant for
frequencies well below and above the relaxation frequency
$\nu_{\mathrm{relax}}=1/(2\pi\tau)$, respectively. This frequency is
characterized by an inflection point in the frequency dependence of
the dielectric constant and a peak in the dielectric loss. If taking
into account an additional dc-conductivity contribution, the
conductivity shows a steplike increase close to
$\nu_{\mathrm{relax}}$. The Debye theory assumes that all entities
do relax with the same relaxation time $\tau$. In reality, a
distribution of relaxation times often leads to a considerable
smearing out of the spectral features \cite{Sillescu1999,
Ediger2000}. Those can be described, e.g., by the Havriliak-Negami
formula, which is an empirical extension of the Debye formula by the
additional parameters $\alpha$ and $\beta$ \cite{Havriliak1966,
Havriliak1967}:

\begin{equation}
\label{hn}
\varepsilon^{*}(\nu)=\varepsilon_{\mathrm{\infty}}+\frac{\Delta\varepsilon}{\left[1+(\mathrm{i}2\pi\nu\tau)^{1-\alpha}\right]^{\beta}}\:
\end{equation}

\noindent Special cases of this formula are the Cole-Cole formula
\cite{Cole1941} with $0\leq\alpha<1$ and $\beta=1$ and the
Cole-Davidson formula \cite{Davidson1950, Cole1952} with $\alpha=0$
and $0<\beta\leq1$.

In most materials with intrinsic relaxations, the inevitable dc
conductivity $\sigma_{\mathrm{dc}}$ arising from ionic or electronic
charge transport cannot be neglected. Usually it leads to a $1/\nu$
divergence in the loss at frequencies below the loss peak and can be
taken into account by including a further additive term
$\varepsilon''_{\mathrm{dc}}=\sigma_{\mathrm{dc}}/(\varepsilon_{0}\omega)$
in Eq. \ref{hn} ($\varepsilon_{0}$ denotes the permittivity of free
space, $\omega$ is the circular frequency).

\subsection{Maxwell-Wagner Relaxations}
\label{Mw}

The $\beta$-dispersion in biological matter is commonly accepted to
be of Maxwell-Wagner type \cite{Schwan1957a, Schwan1983, Pethig1979,
Foster1989, Foster1995}. As shown by Maxwell and Wagner
\cite{Maxwell1873, Wagner1914}, strong dispersive effects mimicking
those of intrinsic dipolar relaxations can arise in samples composed
of two or more regions with different electrical properties (e.g.,
plasma, cytoplasma, and cell membranes in the case of RBCs). It
should be noted that this dispersion can be completely understood
from the heterogeneity of the investigated samples without invoking
any frequency-dependent microscopic processes within the involved
dielectric materials. If one of the regions in the sample is of
interfacial type and relatively insulating, e.g., an insulating
surface layer \cite{Lunkenheimer2002} or the membranes of biological
cells \cite{Schwan1957a,Schwan1983}, very high apparent values of
the dielectric constant are detected at low frequencies. A
straightforward approach for understanding the dielectric behavior
of heterogeneous systems is an equivalent-circuit analysis. Here any
interfacial layer can be modeled by a parallel RC element with the
resistance $R$ and capacitance $C$ of the interfacial element much
higher than the corresponding bulk values \cite{Lunkenheimer2002}.
This leads to a relaxation spectrum where the low-frequency
capacitance and conductance are dominated by the interface. For the
calculation of $\varepsilon'(\nu)$ from the measured capacitance,
usually the overall geometry of the sample is used instead of that
of the thin interfacial layer (i.e., the assumed $C_{0}$ is much
smaller than that of the layer). Thus, an artificially high
dielectric constant is detected at low frequencies. At high
frequencies, the interface capacitor becomes shorted and the bulk
properties are detected. This leads to the steplike decrease of
$\varepsilon'(\nu)$ and increase of $\sigma'(\nu)$ with increasing
frequency, typical for relaxational behavior. The increase of
$\sigma'(\nu)$ arises from the fact that the bulk conductance
usually is much higher than the interface conductance (the cell
membrane in the case of RBCs), the latter being shorted by the
interface capacitor at high frequencies.

Instead of an equivalent-circuit analysis \cite{Lunkenheimer2002},
for biological matter it is common practice to treat the
$\beta$-relaxation analogous to an intrinsic relaxation process,
i.e., to fit it with the Debye equation or its extensions (Eqs.
\ref{Db} and \ref{hn}). A variety of models have been developed to
connect the obtained fitting parameters with the intrinsic
dielectric properties of the different regions of the samples (see
section \ref{cellmod}) and a lot of modeling work of experimental
data was performed \cite{Schwan1957a, Fricke1923, Fricke1924,
Fricke1925a, Fricke1925b, Bruggemann1935, Hayashi2008, Pauly1959,
Hanai1960a, Looyenga1965, Asami2002, Hanai1961, Hanai1968,
Katsumoto2008, Asami1999, Gheorghiu1999, DiBiasio2005, Merla2006,
Sudsiri2007, Hayashi2009}.

\subsection{Electrode Polarization}
\label{BE}

Blood exhibits strong ionic conductivity. At low frequencies the
ions arrive at the metallic electrodes and accumulate in thin layers
immediately below the sample surface \cite{Oncley1942, Schwan1968,
MacDonald1987, Bordi2001}. The frequency, below which this effect
sets in, critically depends on electrode distance and ionic mobility
and in biological matter typically is located in the kHz - MHz
region. These insulating layers represent large capacitors leading
to an apparent increase of $\varepsilon'(\nu)$ and decrease of
$\sigma'(\nu)$ at low frequencies, quite similar to the
Maxwell-Wagner effects discussed above. These non-intrinsic
contributions can hamper the unequivocal detection of the parameters
of the $\beta$-relaxation. Various experimental techniques have been
applied to avoid the influence of electrode polarization (see, e.g.,
refs. \citenum{Bothwell1956, Lisin1996, Schwan1968}). An alternative
way is the exact modeling of these non-intrinsic contributions. The
most common models are a parallel RC circuit or a so-called
constant-phase element, both connected in series to the bulk sample
\cite{Schwan1968, MacDonald1987, Bordi2001}.

A parallel RC circuit corresponds to an additional impedance

\begin{equation}
\label{RC}
Z_{\mathrm{RC}}^{*}(\nu)=\frac{R_{\mathrm{RC}}}{1+\mathrm{i}2\pi\nu
R_{\mathrm{RC}}C_{\mathrm{RC}}},
\end{equation}

\noindent which has to be added to the bulk impedance.
$R_{\mathrm{RC}}$ and $C_{\mathrm{RC}}$ are the resistance and
capacitance of the insulating layers, respectively. From the
resulting total impedance, the total capacitance and conductance
(and thus $\varepsilon'(\nu)$ and $\sigma'(\nu)$) can be calculated
resulting in a behavior equivalent to a Debye-relaxation (this
scenario corresponds to a conventional Maxwell-Wagner relaxation).
Alternatively a "constant phase element", which is an empirical
impedance function, given by
$Z_{\mathrm{CPE}}=A(\mathrm{i\omega})^{-\alpha}$ (refs
\citenum{MacDonald1987, Bordi2001}), can be used. When defining
$\tau_{\mathrm{RC}}=R_{\mathrm{RC}}C_{\mathrm{RC}}$, Eq. \ref{RC}
formally has the same mathematical structure as Eq. \ref{Db}. Thus,
in analogy to Eq. \ref{hn} a distributed RC-circuit can be introduced
by writing

\begin{equation}
\label{RCverteilt}
Z_{\mathrm{RC}}^{*}(\nu)=\frac{R_{\mathrm{RC}}}{\left[1+(\mathrm{i}2\pi\nu\tau_{\mathrm{RC}})^{1-\alpha}\right]^{\beta}}\:.
\end{equation}

\noindent We want to emphasize, that in contrast to Eq. \ref{RC},
which leads to a frequency dependence identical to that of a Debye
relaxation, an equivalent-circuit evaluation using Eq.
\ref{RCverteilt} does not lead to fit curves identical to those of a
Havriliak-Negami relaxation: For the latter case, the relaxation
time $\tau$ in Eq. \ref{Db} is assumed to be distributed. In the
equivalent-circuit case, the corresponding quantity, determining,
e.g., the loss peak position, is $\tau=R_{b}C_{RC}$ with $R_{b}$ the
bulk resistance \cite{Lunkenheimer2002}. However, the distributed
quantity in the equivalent-circuit case is $\tau_{RC}=R_{RC}C_{RC}$,
thus leading to different curve shapes.



\subsection{Temperature Dependence}

The fitting of relaxation spectra directly provides the relaxation
times ($\tau$), the width parameters ($\alpha$ and $\beta$), the
relaxation strengths ($\Delta\varepsilon$), the dielectric constant
for $\nu\longrightarrow\infty$ ($\varepsilon_{\infty}$), and the dc
conductivity ($\sigma_{\mathrm{dc}}$) (see section \ref{Intrinsic}). For the temperature dependence of  $\tau$ and
$\sigma_{\mathrm{dc}}$, thermally activated behavior

\begin{equation}
\label{tautemp}
\tau=\tau_{0}\exp\left(\frac{E_{\tau}}{k_{\mathrm{B}}T}\right)
\end{equation}

\noindent and

\begin{equation}
\label{sdc}
\sigma_{\mathrm{dc}}=\frac{\sigma_{0}}{T}\exp\left(-\frac{E_{\sigma}}{k_{\mathrm{B}}T}\right),
\end{equation}

\noindent can be assumed. $\sigma_{0}$ is a prefactor. $E_{\tau}$
and $E_{\sigma}$ denote the hindering barriers for the relaxational
process and the diffusion of the charge carriers (i.e., dissolved
ions of the plasma in the present case), respectively. $\tau_{0}$ is
an inverse attempt frequency, often assumed to be of the order of a
typical phonon frequency. Equation \ref{sdc}, with the extra $1/T$
term, is derived by considering the difference of the forward and
backward hopping probabilities of an ion between two sites in a
potential that becomes asymmetric due to the external field
\cite{Zarzycki1991}. Here the field drives the ionic motion. In
contrast, for dielectric relaxations, within the framework of the
fluctuation-dissipation theorem it is assumed that the dielectric
measurement is sensitive to reorientational fluctuations, which are
present even without field. Thus, $\tau$ in Eq. \ref{tautemp} is
proportional to the inverse of the reorientation probability of a
molecule experiencing a hindering barrier, which is just given by
the exponential term.

The temperature dependence of the dielectric strength of dipolar
relaxation mechanisms often can be characterized by the Curie law
\cite{Debye1912, Onsager1936}:

\begin{equation}
\label{Curie} \Delta\varepsilon=\frac{C}{T}\
\end{equation}

\noindent Deviations from the Curie law are usually thought to
signify dipole-dipole interactions.

For most dielectric materials, the broadening of the loss peak
diminishes with increasing temperature, i.e., $\alpha\rightarrow0$
and $\beta\rightarrow1$ for high temperatures. This finding can be
ascribed to the fast thermal fluctuations, which cause every
relaxing entity "seeing" the same environment
\cite{Lunkenheimer2000}, thus leading to an identical relaxation
time for each entity, which implies Debye-like behavior.

\subsection{Cell Models} \label{cellmod}

Two commonly employed models for the description of the
$\beta$-dispersion of cell suspensions and tissue will be introduced
now and applied to the experimental data in section \ref{app}. Both
models are claimed to be applicable to high cell concentrations and
thus are especially suited for the samples of the present work.
Based on the Maxwell-Wagner model \cite{Maxwell1873, Wagner1914},
the Pauly and Schwan takes into account the membranes of cells
\cite{Pauly1959, Foster1995}. Using appropriate approximations
(e.g., a negligible membrane conductance) some simple relations are
derived:

\begin{equation}
\label{deltaeps}
\Delta\varepsilon_{\beta}=\frac{9prC_{\mathrm{m}}}{4\varepsilon_{0}\cdot
(1+p/2)^2}
\end{equation}

\noindent and

\begin{equation}
\label{au}
\sigma_{\mathrm{dc\beta}}=\frac{1-p}{1+p/2}\cdot\sigma_{\mathrm{a}}
\end{equation}


\noindent Equation \ref{deltaeps} allows the calculation of the
membrane capacitance per area unit, $C_{\mathrm{m}}$, from the
relaxation strength of the $\beta$-dispersion,
$\Delta\varepsilon_{\beta}$, and the volume fraction $p$ and radius
$r$ of the suspended particles. Via Eq. \ref{au}, the conductivity of
the suspending medium (plasma in the case of blood),
$\sigma_{\mathrm{a}}$, can be determined from the volume fraction
and the measured dc conductivity of the suspension, i.e. the
limiting low frequency conductivity of the $\beta$-dispersion,
$\sigma_{\mathrm{dc\beta}}$. Moreover, resolving the expression for
the $\beta$-relaxation time $\tau_{\mathrm{\beta}}$
\cite{Pauly1959},

\begin{equation}
\label{taubet}
\tau_{\beta}=rC_{\mathrm{m}}\frac{2\sigma_{\mathrm{a}}+\sigma_{\mathrm{i}}-p(\sigma_{\mathrm{i}}-\sigma_{\mathrm{a}})}{\sigma_{\mathrm{i}}\sigma_{\mathrm{a}}(2+p)},
\end{equation}

\noindent one can approximate the conductivity of the cell interior
(cytoplasma), $\sigma_{\mathrm{i}}$ by


\begin{equation}
\label{sp}
\sigma_{\mathrm{i}}=\frac{\sigma_{\mathrm{a}}rC_{\mathrm{m}}(2+p)}{\sigma_{\mathrm{a}}\tau_{\beta}(2+p)-(1-p)rC_{\mathrm{m}}}\:.
\end{equation}

\noindent An alternative access to $\sigma_{\mathrm{i}}$ is provided
by the following relation \cite{Pauly1959}:

\begin{equation}
\label{spsu}
\sigma_{\mathrm{i}}=\frac{2\sigma_{\mathrm{a}}^{2}(1-p)-\sigma_{\mathrm{a}}\sigma_{\infty\beta}(2+p)}{\sigma_{\infty\beta}(1-p)-\sigma_{\mathrm{a}}(1+2p)}
\end{equation}

\noindent Here $\sigma_{\infty\beta}$ denotes the high-frequency
plateau of the step in $\sigma'(\nu)$. Also the dielectric constant
of the cytoplasma, $\varepsilon_{\mathrm{i}}$, can be determined
\cite{Pauly1959}:

\begin{equation}
\label{sp2}
\varepsilon_{\mathrm{i}}=\frac{2\varepsilon_{\mathrm{a}}^{2}(1-p)-\varepsilon_{\mathrm{a}}\varepsilon_{\infty\beta}(2+p)}{\varepsilon_{\infty\beta}(1-p)-\varepsilon_{\mathrm{a}}(1+2p)}
\end{equation}

\noindent $\varepsilon_{\mathrm{\infty\beta}}$ is the limiting
high-frequency dielectric constant of the $\beta$-relaxation (cf.
$\varepsilon_{\mathrm{\infty}}$ in Eq. \ref{Db}) and
$\varepsilon_{\mathrm{a}}$ is the dielectric constant of the
suspending medium (plasma).

The Pauly-Schwan model includes correction factors for high
concentrations (e.g., the $1+p/2$ factor in eq. \ref{deltaeps}) and
is claimed to be valid for all values of $p$ (see, e.g., ref.
\cite{Foster1995}). Another model, especially developed for highly
concentrated suspensions, is the one by Bruggemann
\cite{Bruggemann1935} and Hanai \cite{Hanai1960a, Hanai1961,
Hanai1968} taking into account the polarization of particles in the
presence of neighboring ones. The model leads to the equations

\begin{equation}
\label{hb1}
\varepsilon_{\mathrm{p}}=\frac{\varepsilon_{\mathrm{a}}(1-p)-\varepsilon_{\mathrm{\infty\beta}}k}{1-p-k},\hspace{0.5cm}
k=\left(\frac{\varepsilon_{\mathrm{a}}}{\varepsilon_{\mathrm{\infty\beta}}}\right)^{1/3}\hspace{0.5cm}
\end{equation}

\noindent and

\begin{equation}
\label{hb2}
\sigma_{\mathrm{p}}=\frac{\sigma_{\mathrm{a}}(1-p)-\sigma_{\mathrm{dc\beta}}k}{1-p-k},\hspace{0.5cm}
k=\left(\frac{\sigma_{\mathrm{a}}}{\sigma_{\mathrm{dc\beta}}}\right)^{1/3}
\end{equation}

\noindent for the dielectric properties $\varepsilon_{\mathrm{p}}$
and $\sigma_{\mathrm{p}}$ of the particles. It should be mentioned
that this model assumes homogeneous particles (i.e. without shell)
and Eqs. \ref{hb1} and \ref{hb2} can be considered providing average
values of the whole cell only.

One should note that the exact solution of the dielectric theory of
suspensions of ellipsoidal particles leads to the prediction of six
separate relaxation processes, namely two per ellipsoid axis,
arising from the Maxwell-Wagner relaxation of the shell and of the
particle interior \cite{Asami2002}. For spheroids, i.e. ellipsoids
with two equal semi-diameters, which may be a good approximation of
RBCs, still four relaxations are expected. Usually in the
application of the Maxwell-Wagner model to cell suspensions,
including the above treated models, various reasonable
approximations are made (e.g., that the membrane thickness is much
smaller than the cell radius) that lead to the prediction of a
single relaxation only.



\section{Materials and methods}
\label{Materials}

To determine the complex dielectric permittivity and conductivity in
a broad frequency range (from 1~Hz to 40~GHz), different measurement
techniques were combined \cite{Schneider2001}. In the frequency
range 1~Hz~-~10~MHz, high precision measurements were performed by
means of a Novocontrol Alpha-A Analyzer. This frequency response
analyzer directly measures the sample voltage and the sample current
by the use of lock-in technique. The ac voltage is applied to a
parallel-plate capacitor made of platinum containing the sample
material (diameter 5~mm, plate distance 0.6~mm). In our earlier
measurements of various materials, platinum was found to minimize
contributions from dissolved ionic impurities arising from the
electrode material. The capacitor is mounted into a N$_{2}$-gas
cryostat (Novocontrol Quatro) for temperature-dependent
measurements. For the measurements in the frequency range
1~MHz~-~3~GHz a coaxial reflection method was used employing the
Agilent Impedance/Material Analyzer E4991A. Here the sample, again
placed in the same parallel-plate capacitor, is connected to the end
of a specially designed coaxial line, thereby bridging inner and
outer conductor \cite{Boehmer1989}. For additional measurements
between 40~Hz-110~MHz, the autobalance bridge Agilent 4294A was
used. Its measurement range overlaps  with that of the other
devices. In all the measurements described above, the applied ac
voltage was 0.1~V. The Agilent "Dielectric Probe Kit" 85070E using
the so called "performance probe" with an E8363B PNA Series Network
Analyzer covered the high frequency range from 100~MHz to 40~GHz. It
uses a so-called open-end coaxial reflection technique, where the
end of a coaxial line is immersed into the sample liquid. The
applied ac voltage was 32~mV. Calibration was performed with the
standards Open, Short, and Water. As any contributions from
parasitic elements are excluded by this technique, the obtained
absolute values were used to correct the results obtained with the
low-frequency techniques, discussed above, for contributions from
stray capacitance.

Blood samples from a healthy person were taken at the hospital
"Klinikum Augsburg". All blood samples were taken from the same
person and various vital parameters of the blood samples were
checked. All samples were taken before a meal, at the same time of
the day. We did not find any significant difference in the
measurement results obtained on samples taken at different days. To
avoid clotting, the samples were prepared with EDTA
(ethylenediaminetetraacetic acid). The influence of different
coagulation inhibitors on the dielectric properties were tested and
found to be insignificant. Besides the whole blood, which was
measured as taken from the body, four other samples with different
hematocrit values ($Hct$) were prepared. $Hct$ is given by the ratio
of volume fraction of the corpuscles (erythrocytes, leukocytes, and
thrombocytes) of the blood and the total volume. The whole blood
used in the present work was found to have $Hct=0.39$. After
centrifugation of the whole blood, the corpuscular parts could be
separated from the plasma by pipetting. By remixing with the plasma
obtained in this way, four additional samples were prepared: plasma
($Hct<5$) and blood with $Hct=0.23$, 0.57, and 0.86. The exact $Hct$
values were determined by taking hemograms with a Beckman-Coulter
Hematology Analyzer at the hospital. For each measurement run with
the different devices fresh blood samples were used.

One may suspect that sedimentation of RBCs and other cells during
the temperature-dependent measurement runs could influence the
measurements. Covering the whole investigated temperature range with
the different measurement devices did take about 2-3 hour only. In
the high-frequency measurements with the "open-end" coaxial
technique, where relatively large amounts of material (about 25~ml)
held in transparent test tubes were used, visible inspection
revealed no indications of sedimentation even after much longer
time. Nevertheless the sample material was thoroughly stirred before
each frequency sweep at the different temperatures. The results did
well match those at lower frequencies, where small sample amounts of
about 0.01~ml contained in a platinum capacitor were used. With this
capacitor, up to three separate temperature-dependent measurement
runs were performed, using different devices and partly using
different thermal histories. All the results did match very well.
Finally, the measurements usually were done by first cooling the
sample from room temperature and subsequently heating it up to the
highest temperatures. The results from the cooling and heating runs,
which were done at differently aged samples, did always agree within
experimental resolution.

The integrity of the erythrocytes was retained during most of the
dielectric measurements, which was checked by a comparison of the
room temperature results before and after cooling or heating.
However, as expected this was no longer the case when the samples
were subjected to the highest temperatures investigated, extending
up to 330~K. Therefore these measurements were performed at the end
of each temperature-dependent measurement run.

If not otherwise indicated, the experimental data points shown in
the plots of the present work have errors that are not exceeding the
size of the symbols.

\section{Results and Discussion}
\label{results}

\subsection{Broadband Spectra}

\begin{figure}[h]
\centering
\includegraphics[width=8cm]{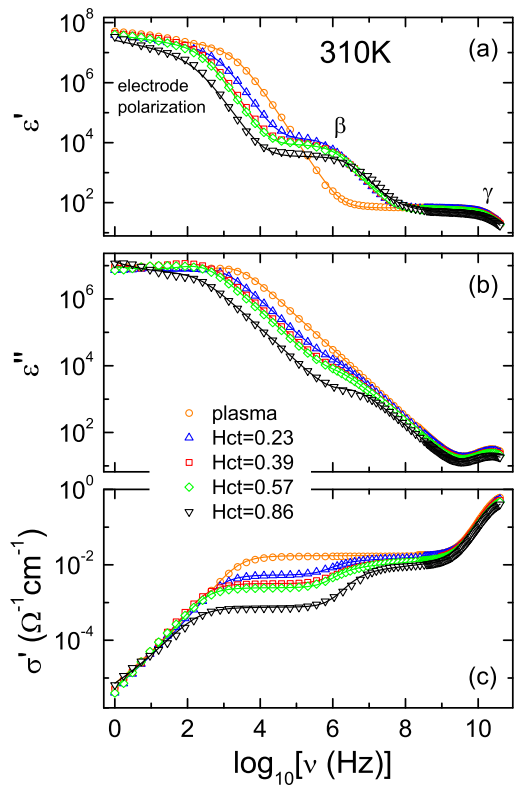}
\caption[Debye-Polarisation]{\label{fig:all} (a) Dielectric
constant, (b) dielectric loss, and (c) real part of the conductivity
of whole blood, blood plasma ($Hct=0.39$), and blood samples with
different hematocrit values as function of frequency, measured at
body temperature (310~K). The lines are fits assuming a distributed
RC equivalent circuit to account for the electrodes and, for samples
with $Hct\geq0.23$, two Cole-Cole functions for the $\beta$- and
$\gamma$-relaxation. For the plasma data, a single Cole-Cole
function was used instead.}
\end{figure}

Figure \ref{fig:all} shows the broadband spectra of the different
samples at body temperature ($\approx310$~K), extending from 1~Hz to
40~GHz. 

The dielectric constant of the blood plasma (Fig.
\ref{fig:all}(a), orange circles) reveals a low-frequency plateau
between 1 and 100~Hz, followed by a steplike decrease towards higher
frequencies that passes into another plateau between about 1~MHz and
10~GHz. The behavior below about 1 MHz can be ascribed to electrode
polarization (see section \ref{alpha} for a detailed discussion). At
frequencies beyond about 1~GHz a further decrease of
$\varepsilon'(\nu)$ indicates the beginning $\gamma$-dispersion
arising from the reorientational motion of the water molecules (see
section \ref{gamma}). $\varepsilon''(\nu)$ shows a plateau at low
frequencies, followed by a strong decrease above about 300~Hz and
the $\gamma$-relaxation peak at ca. 20~GHz. Accordingly, the
conductivity $\sigma'(\nu)$ exhibits a strong increase at low
frequencies, followed by a plateau between 1~kHz and 1~GHz. At
$\nu>1$~GHz another strong increase at the end of the measured
spectrum shows up, again corresponding to the $\gamma$-relaxation.

Just as the plasma, the dielectric spectra of the other samples also
show a $\gamma$-relaxation and a electrode-polarization
contribution, the latter leading to a strong increase of
$\varepsilon'(\nu)$ below about 10 - 100~kHz and decrease of
$\sigma'(\nu)$ below about 1 - 10~kHz. However, between these two
features, a further process shows up, the well-known
$\beta$-dispersion, located at about 1 - 100~MHz in
$\varepsilon'(\nu)$. It is caused by the Maxwell-Wagner relaxation
arising from the heterogeneity of the solute/cell system (see
section \ref{Mw}). It is evidenced by a steplike decrease in the
dielectric constant at about 1 - 100~MHz, an s-shaped bend in the
decrease of the dielectric loss around 1~MHz, and a steplike
increase in the conductivity around 1~MHz. Comparing the different
samples, it becomes evident that the absolute values of
$\varepsilon'(\nu)$, $\varepsilon''(\nu)$, and $\sigma'(\nu)$
decrease with increasing hematocrit value over almost the whole
frequency range. A detailed analysis of the $\beta$-dispersion will
be provided in section \ref{beta}. The $\delta$-dispersion, which is
supposed to be located in the frequency range between the $\beta$-
and $\gamma$-relaxation cannot be detected on this scale and will be
treated in section \ref{further} below.

In the present work the complete broadband spectra are fitted by
combining several relaxational dispersions and the
electrode-polarization contribution. In addition, the dc
conductivity has to be included in the fitting routine, since the
blood plasma contains free ions that contribute to the conductivity.
The lines shown in Fig. \ref{fig:all} are fits with the sum of two
relaxational dispersions described by Eq. \ref{hn} (with $\beta=1$)
and the dc-conductivity, which are assumed to be connected in series
to the electrode impedance given by Eq. \ref{RCverteilt} (with
$\beta=1$). The fits were simultaneously performed for real and
imaginary part of the permittivity. A qualitative inspection of Fig.
\ref{fig:all} reveals that an excellent match of the experimental
spectra could be achieved in this way, which also is the case for
the other temperatures investigated. In the following sections, we
discuss the different contributions to the spectra and the resulting
relevant fit parameters in detail. To serve for SAR calculations and
medical purposes, the fit curves for all temperatures and hematocrit
values investigated in the present work are provided in electronic
form.

\subsection{Electrode polarization and $\alpha$-Dispersion}
\label{alpha}

\begin{figure}[h]
\centering
\includegraphics[width=8cm]{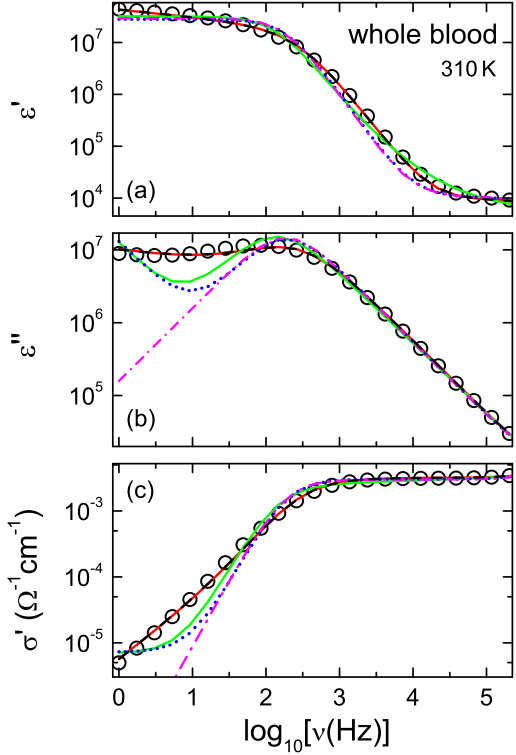}
\caption[Debye-Polarisation]{\label{fig:alphaRCCC}$\varepsilon'(\nu)$
(a), $\varepsilon''(\nu)$ (b), and $\sigma'(\nu)$ (c) of whole blood
($Hct=0.39$) at 310~K and low frequencies (circles). The lines are
fits using different functions: RC equivalent circuit (Eq. \ref{RC};
dash dotted magenta line), distributed RC equivalent circuit (Eq.
\ref{RCverteilt}; red line), Debye function with additional
dc-conductivity contribution (dotted blue line), Cole-Cole function,
also with dc contribution (green line), and constant phase element
(black dashed line).}
\end{figure}

To examine the contributions from electrode polarization to the
spectra in more detail, as an example in Fig. \ref{fig:alphaRCCC}
the spectra of whole blood at 310~K are shown in the low frequency
range, 1~Hz - 200~kHz. As demonstrated by the dash-dotted magenta
line, the observed relaxationlike feature cannot be satisfactorily
fitted by assuming a simple RC equivalent circuit (Eq. \ref{RC}) in
series to the bulk sample, which leads to a symmetric loss peak just
like an intrinsic Debye relaxation (see section \ref{BE}). The
deviations are strongest in $\varepsilon''$ below the peak
frequency, where the measured $\varepsilon''(\nu)$ obviously varies
far too weakly with frequency to be describable by the strongly
increasing fit curve. Using a Debye relaxation function (Eq.
\ref{Db}) with an additional conductivity contribution
$\sigma'_{\mathrm{dc}}$, which leads to a minimum in
$\varepsilon''(\nu)$ below the peak frequency, also cannot account
for the experimental data (dotted blue line). Replacing the Debye by
a Cole-Cole function (Eq. \ref{hn} with $\beta=1$; green line) only
leads to a marginal improvement of the fits. Instead, a distributed
RC equivalent circuit as described by Eq. \ref{RCverteilt} with
$\beta=1$ but $\alpha\neq0$ (Cole-Cole case) was found to provide
very accurate fits of the measured spectra (solid red line). The
same can be said for the other blood samples with different
hematocrit values investigated in the present work. A constant phase
element in series with the bulk also very well accounts for the
experimental data (black dashed line). However, in our further
analysis we decided to use the distributed RC circuit instead
because its parameters seem to have more physical signification than
those of the constant phase element.

\begin{figure}[h]
\centering
\includegraphics[width=8cm]{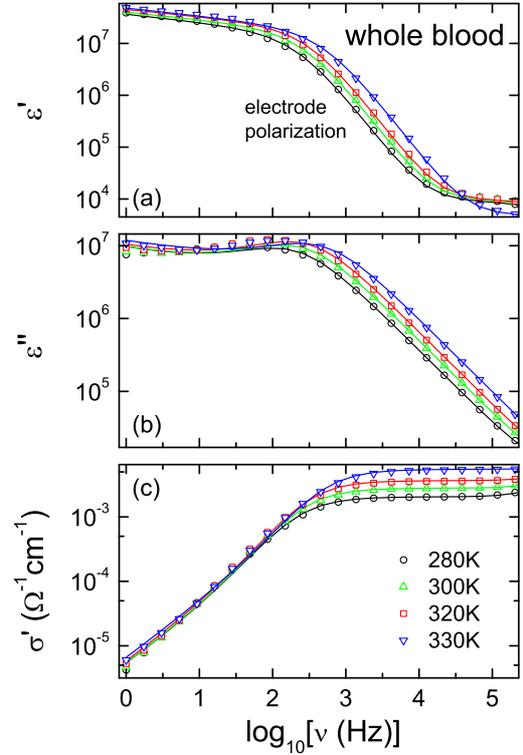}
\caption[Debye-Polarisation]{\label{fig:alphaVoll}$\varepsilon'(\nu)$
(a), $\varepsilon''(\nu)$ (b), and $\sigma'(\nu)$ (c) of whole blood
($Hct=0.39$) in the low-frequency regime dominated by electrode
effects shown for selected temperatures (symbols). The lines
represent fit curves as in Fig. \ref{fig:all} using a distributed RC
equivalent circuit for the description of the electrode
polarization.}
\end{figure}

Good fits with this approach can also be achieved for the results
obtained at different temperatures. As an example, Fig.
\ref{fig:alphaVoll} shows the dielectric quantities of whole blood
in the frequency range, dominated by electrode polarization, for
selected temperatures. The lines represent the fits of the complete
broadband spectra (cf. Fig. \ref{fig:all}) where a distributed RC
circuit, Eq. \ref{RCverteilt}, was used for the description of the
low-frequency data. In all cases the agreement of fits and
experimental curves are excellent. The onset of the electrode
effects, i.e. the increase of $\varepsilon'$ and the decrease of
$\sigma'$ when lowering the frequency, is found to shift to lower
frequencies with decreasing temperature. This can be ascribed to the
reduced mobility of the ionic charge carriers at low temperatures,
which thus arrive at the electrodes for smaller frequencies only. As
revealed by Fig. \ref{fig:all}, for increasing hematocrit value the
onset of the electrode effects shifts to lower frequencies, too.
This is in accord with the well-known increase of the viscosity of
blood with increasing hematocrit value, corresponding to a reduction
of ion mobility.

The fits reveal a width parameter $\alpha$ close to 0.15 and nearly
independent of temperature and hematocrit value (not shown), except
for $Hct=0.86$, where $\alpha\approx0.20$ is found, slightly
increasing with temperature. $\alpha$ characterizes the distribution
of relaxation times of the RC equivalent circuit that describes the
electrode polarization (see section \ref{BE}). The deviations from
Debye behavior may be explained, e.g., by the surface roughness of
the electrodes \cite{Nyikos1985, Liu1985a, Pajkossy1986}. The fits
reveal electrode capacitances $C_{\mathrm{RC}}$ of the order of
10~$\mu$F. $C_{\mathrm{RC}}$ is found to be only weakly temperature
dependent and it shows a tendency to decrease with increasing
hematocrit value. This can be ascribed to the mentioned reduction of
the ionic mobility leading to a less effective formation of the
insulating electrode layers. The fits do not reveal reliable
information on the electrode resistance as no clear low-frequency
plateau in $\sigma'(\nu)$ is seen (cf. Fig. \ref{fig:all} and Fig.
\ref{fig:alphaVoll}).

From the presented results, it is clear that the electrode
polarization is the dominant effect in the low frequency spectrum of
blood. The equivalent-circuit description in terms of a distributed
RC circuit provides nearly perfect fits of the experimental data.
The typical deviations between fit and measured data are around
10$\%$ or less, which is negligible compared to the many decades the
dielectric quantities vary with frequency. Thus the presence of an
additional contribution from a possible $\alpha$-relaxation seems
unlikely but due to the mentioned deviations, a weak
$\alpha$-relaxation cannot be fully excluded. However, it should be
noted that in earlier investigations also no indications for an
$\alpha$-relaxation in blood were found \cite{Bothwell1956}.

\subsection{$\beta$-Dispersion}
\label{beta}

\subsubsection{Phenomenological Evaluation}
\label{sub:beta}

\begin{figure}[h]
\centering
\includegraphics[width=8cm]{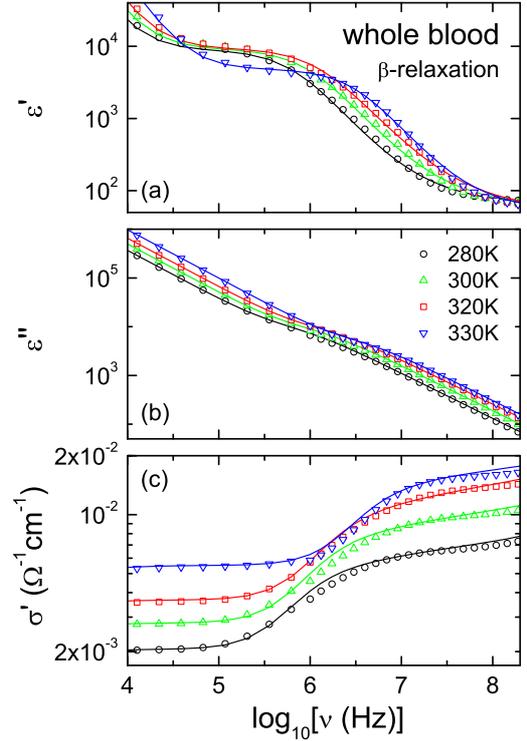}
\caption[Debye-Polarisation]{\label{betaVollblut}$\varepsilon'(\nu)$
(a), $\varepsilon''(\nu)$ (b), and $\sigma'(\nu)$ (c) of whole blood
($Hct=0.39$) in the $\beta$-dispersion region for selected
temperatures. The lines represent fit curves as in Fig.
\ref{fig:all} using the Cole-Cole function for the description of
the $\beta$-relaxation.}
\end{figure}

Figure \ref{betaVollblut} shows the spectra of whole blood in the
frequency range of the $\beta$-dispersion (10~kHz to 200~MHz) at
different temperatures. Except for the plasma, not containing any
RBC's that would cause a $\beta$-process, all samples show a similar
relaxational behavior and temperature dependence in this frequency
regime (see Fig. \ref{fig:all}). Just as for intrinsic relaxations,
the permittivity curves in the $\beta$-dispersion regime shift to
higher frequencies with increasing temperature. The lines in Fig.
\ref{betaVollblut} again represent the fits of the complete
broadband spectra (cf. Fig. \ref{fig:all}) using a Cole-Cole
function (Eq. \ref{hn} with $\beta=1$) for the $\beta$-relaxation.
As discussed in detail in section \ref{Mw}, the $\beta$-relaxation
is generated by the heterogeneity of the sample material, which is
composed of plasma and RBC's. $\varepsilon'(\nu)$ exhibits the
typical steplike decrease with increasing frequency (Fig.
\ref{betaVollblut}(a)). Its additional increase towards the lowest
frequencies observed in Fig. \ref{betaVollblut}(a) corresponds to
the onset of the electrode-polarization effects (cf. Fig.
\ref{fig:alphaVoll}), well taken into account in the fits by the
distributed RC equivalent circuit (see section \ref{alpha}). In
$\varepsilon''(\nu)$ a loss peak is expected but only its
high-frequency flank can be seen (Fig. \ref{betaVollblut}(b)). Its
low-frequency part is superimposed by the strong ionic dc
conductivity, which leads to a contribution
$\varepsilon''_{\mathrm{dc}}=\sigma_{\mathrm{dc}}/(\varepsilon_{0}\omega)$.
Thus, instead of a loss peak, only a slight shoulder at about 3~MHz
is revealed.

The dc conductivity also leads to the low-frequency plateau in
$\sigma'(\nu)$ (Fig. \ref{betaVollblut}(c)) while the shoulders
observed around 3~MHz arise from the relaxation. Via the relation
$\sigma'=\varepsilon_{0}\varepsilon''\omega$, the nearly Debye-like
behavior of the $\beta$-relaxation (implying
$\varepsilon''(\nu)\sim\nu^{-1}$ on the high frequency side of the
peaks) leads to the nearly frequency independent $\sigma'(\nu)$ at
$\nu>10$~MHz. The low- and high-frequency plateaus of $\sigma'$ are
labeled as $\sigma_{\mathrm{dc\beta}}$ and
$\sigma_{\mathrm{\infty\beta}}$, respectively. The steplike increase
of $\sigma'(\nu)$ from $\sigma_{\mathrm{dc\beta}}$ to
$\sigma_{\mathrm{\infty\beta}}$ can be qualitatively understood
assuming a shorting of the cell membrane capacitances at high
frequencies. Thus, at high frequencies the RBC's no longer obstruct
the current path and an enhanced conductivity is detected. Therefore
$\sigma_{\mathrm{\infty\beta}}$ can be regarded as good
approximation of the intrinsic conductivity of the plasma, denoted
as $\sigma_{\mathrm{a}}$ in section \ref{cellmod} (in fact it is a
mixture of plasma and cytoplasma conductivity, which we here assume
to be of not too different magnitude). This is nicely corroborated
by the approximate agreement of this plateau value with the
conductivity of the pure plasma as seen in Fig. \ref{fig:all}(c) for
all $Hct$ values (the small deviations for higher $Hct$ values are
due to the larger volume fraction of cytoplasma having somewhat
lower conductivity; see section \ref{app}). The absolute values of
both conductivity plateaus revealed in Fig. \ref{betaVollblut}(c)
increase with increasing temperature, mirroring the thermally
activated ionic charge transport in the plasma.

\begin{figure}[h]
\centering
\includegraphics[width=8cm]{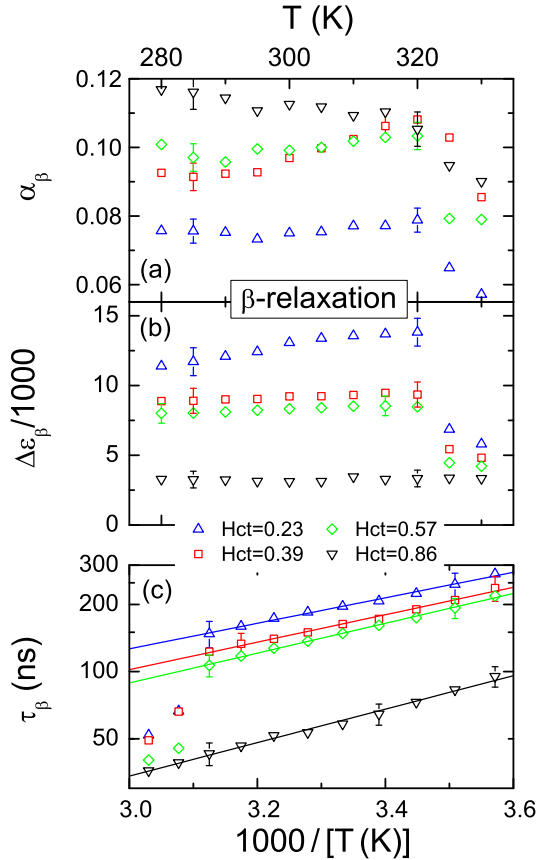} 
\caption[Debye-Polarisation]{\label{betapar} Temperature dependence
of width parameter (a), relaxation strength (b), and relaxation time
(c) as obtained from fits assuming a Cole-Cole-function for the
description of the $\beta$-relaxation (cf Fig. \ref{betaVollblut}).
The lines in the Arrhenius plot of $\tau_{\beta}$ (c) are linear
fits corresponding to thermally activated behavior, Eq.
\ref{tautemp}.}
\end{figure}

As mentioned above, the best fitting results of the
$\beta$-relaxation were achieved by using a Cole-Cole function. Only
for the highest hematocrit values, significant deviations of fits
and experimental data were observed, which will be treated in
section \ref{further}. The temperature dependence of the width
parameter $\alpha_{\beta}$, the relaxation strength
$\Delta\varepsilon_{\beta}$, and the relaxation time $\tau_{\beta}$
obtained from the fits are shown in Fig. \ref{betapar}. The present
fitting of the complete broadband spectra, including the
contributions from electrode polarization and the
$\gamma$-relaxation, minimizes the influence of any additional
processes on the obtained fit parameters.

The width parameter $\alpha$ (Fig. \ref{betapar}(a)), which for
intrinsic relaxations usually is assumed to arise from a
distribution of relaxation times \cite{Sillescu1999, Ediger2000}, is
almost temperature independent. As $\alpha$ assumes rather small
values between 0.07 and 0.11, the $\beta$ process shows nearly
Debye-like behavior. The width parameter increases with increasing
RBC content, i.e., the deviations from the Debye case become
stronger. According to the Pauly-Schwan model (see section
\ref{cellmod}, Eq. \ref{taubet}) the relaxation time $\tau_{\beta}$
depends on the conductivity outside ($\sigma_{\mathrm{a}}$) and
inside of the cell ($\sigma_{\mathrm{i}}$) and on the membrane
capacitance ($C_{\mathrm{m}}$). It is unlikely that the cell
parameters $\sigma_{\mathrm{i}}$ or $C_{\mathrm{m}}$ should be
influenced by the hematocrit value and thus a distribution of the
outer plasma conductivity seems the most likely cause of the
non-Debye behavior. But also an alternative explanation seems
possible: $\alpha\neq0$ implies a shallower high-frequency flank of
the $\beta$-peak. This flank is essentially determined by the
intrinsic plasma conductivity and corresponds to the high-frequency
plateaus seen in Fig. \ref{betaVollblut}(c). Thus $\alpha\neq0$
implies an increase of $\sigma'(\nu)$ with frequency, which is
typical for hopping conductivity as commonly found for ionic charge
transport \cite{Jonscher1983,Elliott1987,Elliott1989,Funke1993}.
Finally, it has to be mentioned that additional relaxations arising
from the other cell types in blood could also influence the observed
$\beta$-relaxation. As RBCs are by far the dominating cell species
(e.g., volume fraction about 45\% vs. $\sim1\%$ of white blood cells),
these contributions can be expected to be small. Nevertheless, it
cannot be excluded that they may contribute to the observed
deviations from Debye behavior.

As revealed by Fig. \ref{betapar}(b), the relaxation strength of the
$\beta$-relaxation is nearly temperature independent. Thus,
according to Eq. \ref{deltaeps}, the membrane capacity also can be
assumed to be temperature independent. The strong drop of
$\Delta\varepsilon_{\beta}$ at $T>320$~K, observed for all samples
except for $Hct=0.86$, is most likely due to the hemolysis of the
RBC's at high temperatures. This assumption is supported by the fact
that no such deviations can be found for the $\gamma$-relaxation
(see section \ref{gamma}), which is independent of the RBC's.
However, it is not clear why the sample with the highest RBC content
(black triangles in Fig. \ref{betapar}(b)) remains unaffected.
Possibly, cell-cell interactions prevent hemolysis at high Hct.

Figure \ref{betapar}(b) also reveals a decrease of the relaxation
strength with increasing content of erythrocytes. This is a rather
surprising result and cannot be explained within the proposed
theories. Especially, it contradicts the increase of
$\Delta\varepsilon_{\beta}$ with $p$ predicted by Eq.
\ref{deltaeps}. In literature, suspensions of erythrocytes and other
cells using non-plasma solvents, like phosphate buffered saline,
quite generally show a continuous increase of
$\Delta\varepsilon_{\beta}$ with $Hct$ \cite{Lisin1996, Hayashi2008,
Fricke1953, Davey1992, Bordi2002}. However, blood samples (i.e. with
the suspending medium being plasma) can show more complex behavior,
especially for higher $Hct$ values \cite{Chelidze2002, Pfutzner1984}
and as shown in ref \citenum{Beving1994a}, substitution of plasma by
some other solute can strongly influence
$\Delta\varepsilon_{\beta}$. In the framework of a simple
equivalent-circuit picture, $\Delta\varepsilon_{\beta}$ is
determined by a complex superposition of the membrane capacitances
of all RBCs. Increasing the number of RBCs could lead to an increase
or decrease of $\Delta\varepsilon_{\beta}$ depending on whether
parallel or series connections of the cell capacitances (relative to
the field direction) are prevailing. The latter seems to be the case
in our blood samples. The reason is unclear but cell aggregation as,
e.g., rouleaux formation may play a role here \cite{Beving1994a,
Pfutzner1984, Pribush1999, Pribush2000}.

According to Eq. \ref{taubet}, the $\beta$-relaxation time should
depend on $\sigma_{\mathrm{i}}$, $\sigma_{\mathrm{a}}$, and
$C_{\mathrm{m}}$. It was shown above that the membrane capacitance
is nearly temperature independent. Thus, the conductivities should
dominate the temperature dependence of $\tau_{\beta}$. Indeed the
Arrhenius representation of Fig. \ref{betapar}(c) reveals that
$\tau_{\beta}(T)$ is in accord with the expected thermally activated
behavior, typical for ionic conductivity. The hindering barriers
$E_{\tau}$, calculated from the slopes in Fig. \ref{betapar}(c) (cf. Eq.
\ref{tautemp}), seem to slightly increase with growing $Hct$ and
barriers varying between 0.11 and 0.15~eV were obtained. However,
due to the rather small temperature region that could be covered in
these biological samples (compared, e.g., to supercooled liquids
\cite{Lunkenheimer2000, Kremer2002}), the significance of these
values should not be overemphasized. In addition, Fig. \ref{betapar}(c)
reveals a decrease of the relaxation times with increasing
hematocrit value. In principle, such a behavior seems to be
consistent with Eq. \ref{taubet} but the observed variation by about
a factor of three is stronger than expected. However, one should be
aware that the relaxation time is directly proportional to
$C_{\mathrm{m}}$ (Eq. \ref{taubet}) while $C_{\mathrm{m}}$ itself is
proportional to $\Delta\varepsilon_{\beta}$ (Eq. \ref{deltaeps}).
Thus it is clear that the observed rather strong $Hct$-dependent
variation of $\tau_{\beta}$ is directly connected to that of
$\Delta\varepsilon_{\beta}$ revealed by Fig. \ref{betapar}(b).


To compare the results on the $\beta$-relaxation parameters
presented above with earlier publications only partly is possible,
because (to the best knowledge of the authors) no such systematic
(temperature and hematocrit dependent) and broadband research on
human blood has been done before. Moreover, the available literature
values deviate quite strongly from each other. For various
erythrocyte suspensions, the reported values of the relaxation time
$\tau_{\beta}$ are, for example, 254~ns ($Hct=0.07$, room
temperature) \cite{Lisin1996}, 29~ns ($Hct=0.30$, $T=298$~K)
\cite{Bordi1990}, or 230~ns ($Hct=0.47$, $T=310$~K) \cite{Bordi1997,
Davey1989}. In blood the following values were found:
$\tau_{\beta}=133$~ns (sheep blood, $T=310$~K) \cite{Gabriel1996b},
$\tau_{\beta}=89-65$~ns (human blood$, T=288-308$~K, $Hct=0.43$)
\cite{Cook1952}, and $\tau_{\beta}=53.1$~ns (bovine blood, room
temperature, $Hct=0.50$) \cite{Schwan1983}. The values in the
present work vary between 35.9~ns ($Hct=0.86$, $T=330$~K) and
274.1~ns ($Hct=0.23$, $T=280$~K). The literature results for the
relaxation strength also show rather strong variation. In ref
\citenum{Bordi1997}, literature values between 1100 and 5000 were
reported. Fricke found, dependent on $Hct$,
$\varepsilon_{\mathrm{s}}\approx\Delta\varepsilon=900-4000$ for dog,
rabbit, and sheep blood \cite{Fricke1953}. For human blood,
Pf\"{u}tzner published values between approximately 2000 and 6000
($Hct=0.10-0.90$) \cite{Pfutzner1984}. In the present work we have
obtained $\Delta\varepsilon\approx3300-13800$ (for $Hct=0.86$,
$T=330$~K and $Hct=0.23$, $T=320$~K, respectively). Even less data
are available for the width parameter $\alpha_{\beta}$, because
often other fitting functions were used. But mostly they are around
0.1 \cite{Gabriel1996b, Bordi1997, Gabrielinet}, similar to the
values in the present work.


\subsubsection{Application of Cell Models}
\label{app}

As mentioned in section \ref{cellmod}, by using appropriate
models it should be possible to determine intrinsic cell parameters
as the membrane capacitance or the conductivity of the cytoplasma
from the parameters of the $\beta$-relaxation. Using the fitting
parameter $\Delta\varepsilon_{\mathrm{\beta}}$ and Eq.
\ref{deltaeps}, the membrane capacitance $C_{\mathrm{m}}$ can be
calculated. As discussed in the previous section, the experimentally
determined dielectric strength decreases with increasing $Hct$, in
contrast to the increase predicted by Eq. \ref{deltaeps}. The use of
Eq. \ref{deltaeps} therefore would imply a strongly $Hct$-dependent
membrane capacitance (see Table \ref{tab:Aus310} for the results at
310~K). This can hardly be interpreted in a physical way. Literature
values vary between 0.17~$\mu$F/$\mathrm{cm}^2$ (ref
\citenum{Bordi1990}) and 3~$\mu$F/$\mathrm{cm}^2$ (ref
\citenum{Beving1994a}). However, most authors assume a membrane
capacity of about 1~$\mu$F/$\mathrm{cm}^2$ \cite{Bordi1997, Fricke1953,
 Davey1992, Ballario1984a, Asami1989}, whereas some report
$Hct$-dependent membrane capacities \cite{Zhao1993, Chelidze2002}.
Possible reasons for the unexpected behavior of
$\Delta\varepsilon_{\mathrm{\beta}}(Hct)$ and thus of
$C_{\mathrm{m}}(Hct)$ were discussed in the previous section. It
seems that Eq. \ref{deltaeps} is not able to account for the
observations in "real" blood samples, in contrast to suspensions of
erythrocytes in common solvents.

\begin{table}[ht]
\centering
\begin{tabular}{ccccc}
\hline
{\rule[-3mm]{0mm}{8mm} $Hct$}& $C_{\mathrm{m}}\,(\frac{\mu\mathrm{F}}{\mathrm{cm^2}}$) & $\sigma_{\mathrm{a}}\,(\frac{10^{-2}}{\Omega \mathrm{cm}})$  & $\sigma_{\mathrm{i}}\,(\frac{10^{-2}}{\Omega \mathrm{cm}})$ &$\varepsilon_{\mathrm{i}}$ \\
\hline
        0.23 &     11 ($\pm 3$) &       0.79 ($\pm 0.15$)&      0.75 ($\pm 0.08$)&     36.9 ($\pm 3.0$)    \\
        0.39 &     4.9 ($\pm 1.4$)&     0.63 ($\pm 0.15$)&      0.50 ($\pm 0.08$) &     41.5 ($\pm 2.5$)    \\
        0.57 &     3.5 ($\pm 1.1$)&       0.74 ($\pm 0.15$)&      0.78 ($\pm 0.09$)&     42.0 ($\pm 2.5$)   \\
        0.86 &     1.2 ($\pm 0.4$) &       0.78 ($\pm 0.15$) &      0.83 ($\pm 0.11$)&     44.7 ($\pm 2.0$)   \\
\hline\end{tabular} \caption{\label{tab:Aus310}Membrane capacitance
$C_{\mathrm{m}}$, plasma conductivity $\sigma_{\mathrm{a}}$,
conductivity of the cell interior $\sigma_{\mathrm{i}}$, and
dielectric constant of the cell interior $\varepsilon_{\mathrm{i}}$
at 310~K as determined from Eqs. \ref{deltaeps}, \ref{au},
\ref{spsu}, and \ref{sp2} respectively.}
\end{table}

Using Eq. \ref{au} should allow for the calculation of the
conductivity of the suspending medium $\sigma_{\mathrm{a}}$ from the
dc conductivity $\sigma_{\mathrm{dc}\beta}$ of the blood samples.
$\sigma_{\mathrm{a}}$ also can be directly determined from the dc
conductivity of the plasma, measured in the present work
($1.7\times10^{-2}~\Omega^{-1}\mathrm{cm}^{-1}$ at 310~K). The
calculated values are provided in Table \ref{tab:Aus310}. While being
nearly $Hct$-independent as expected, they differ from the directly
measured value by a factor of about two. Again cell aggregation
causing a lowering of the experimentally observed
$\sigma_{\mathrm{dc}\beta}$ may explain this finding.

\begin{figure}[h]
\centering
\includegraphics[width=8cm]{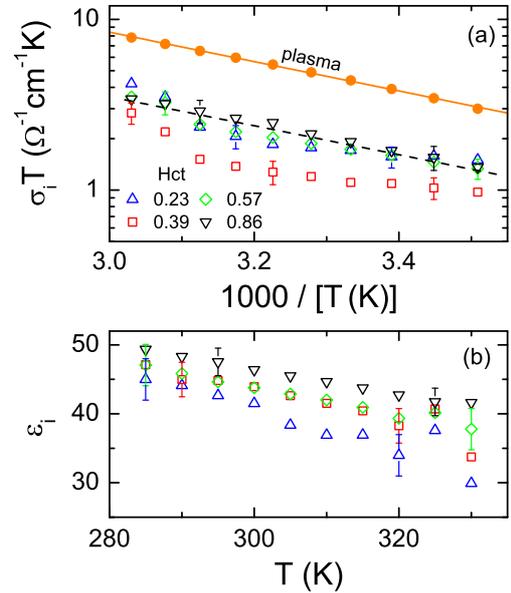}
\caption[Debye-Polarisation]{\label{fig:innen} Temperature dependent
conductivity (Arrhenius plot) (a) and dielectric constant (b) of the
cell interior calculated from Eqs. \ref{spsu} and \ref{sp2},
respectively. For comparison, in (a) also data for pure plasma are
provided. The solid line in (a) indicates approximately linear
behavior implying thermally activated charge transport (Eq.
\ref{sdc}). The dashed line shows an approximate description of the
blood data (except for $Hct=0.39$) using the same energy barrier as
for the plasma.}
\end{figure}

Via Eq. \ref{sp}, the conductivity of the cell interior
$\sigma_{\mathrm{i}}$ can be calculated from $\sigma_{\mathrm{a}}$,
$C_{\mathrm{m}}$, and the measured $\beta$-relaxation times.
However, as $C_{\mathrm{m}}$ determined from Eq. \ref{deltaeps} shows
an unphysical $Hct$ dependence (Table \ref{tab:Aus310}), also an
unreasonable $Hct$ dependence of $\sigma_{\mathrm{a}}$ would result.
An alternative determination of $\sigma_{\mathrm{i}}$ is provided by
Eq. \ref{spsu}. Using the experimentally determined
$\sigma_{\mathrm{a}}$ and $\sigma_{\infty\beta}$ at 310~K we arrive
at the values for $\sigma_{\mathrm{i}}$ listed in Table \ref{tab:Aus310}.
Literature results are distributed around
$0.6\times10^{-2}(\pm0.1)~\Omega^{-1}\mathrm{cm}^{-1}$
\cite{Pauly1966, Krupa1972, Beving1994a, Chelidze2002, Hayashi2008, Bordi2002, Ballario1984a, Asami1989,
Bianco1979}. Deviating
results were reported by Cook \cite{Cook1951, Cook1952} (ca.
$1.0\times10^{-2}~\Omega^{-1}\mathrm{cm}^{-1}$) and Asami
\cite{Asami1980} (ca.
$0.32\times10^{-2}~\Omega^{-1}\mathrm{cm}^{-1}$). Our values for
$\sigma_{\mathrm{i}}$ are by about a factor 2-3 smaller than the
measured conductivity of the plasma,
$\sigma_{\mathrm{a}}\approx1.7\times10^{-2}~\Omega^{-1}\mathrm{cm}^{-1}$.
It seems reasonable that the conductivity of the cytoplasma should
be lowered by the presence of the large hemoglobin molecules and
their bound water shells within the RBCs (about 37\% volume fraction
\cite{Pauly1966}), if compared to the conductivity of the outer
plasma (see following discussion of Fig. \ref{fig:innen} for a
quantitative treatment). Indeed such behavior was found previously
\cite{Krupa1972, Beving1994a, Chelidze2002}. The reported ratios
between about 1.5 and 2.7 are consistent with our findings.
Obviously, Eq. \ref{spsu} is able to provide reasonable estimates for
$\sigma_{\mathrm{i}}$. It is based on the determination of
$\sigma_{\infty\beta}$, which is read off at high frequencies, where
the cell membranes are shorted and thus cell aggregation has no
effect on the results.

In Fig. \ref{fig:innen}(a) the temperature dependence of
$\sigma_{\mathrm{i}}$ is shown in the Arrhenius type of presentation
($\log(\sigma_{\mathrm{i}}T)$ vs $1000/T$) commonly used for ionic
conductivity (cf Eq. \ref{sdc}). For comparison, also the dc
conductivity determined from fits of the spectra of pure blood
plasma (cf. Fig. \ref{fig:all}) is included. In the determination of
$\sigma_{\mathrm{i}}(T)$ via Eq. \ref{spsu}, for
$\sigma_{\mathrm{a}}$ the plasma dc conductivity was used and
$\sigma_{\infty\beta}$ was calculated from the fit parameters of the
$\beta$-relaxation shown in Fig. \ref{betapar}. As the $\beta$-relaxation
shows slight deviations from Debye behavior, $\sigma'(\nu)$ exhibits
a slight increase in the region of its high-frequency plateau (see,
e.g., Fig. \ref{betaVollblut}(c)). As an estimate of
$\sigma_{\infty\beta}$, we used the value of $\sigma'(\nu)$ at a
frequency 1.5 decades above the peak frequency. Except for whole
blood, the temperature dependence of $\sigma_{\mathrm{i}}$ is nearly
independent of $Hct$ and seems to follow thermally activated
behavior (dashed line in Fig. \ref{fig:innen}(a)). The deviations at the
two highest temperatures are directly related to similar problems in
the $\beta$-relaxation parameters (Fig. \ref{betapar}). As discussed in section
\ref{beta}, this may arise from an onset of hemolysis of the RBCs. Interestingly,
the energy barrier of 0.17~eV, deduced from the slope of the
linear fit curve of the plasma data (solid line) is
in good accord with the results on the blood samples (dashed line).
Nevertheless, the absolute values of the conductivity of the
cytoplasma are about a factor of 2-3 lower than those of the plasma.
As mentioned in the previous paragraph, this can be explained by the
presence of the hemoglobin molecules within the cell. For a
quantitative estimate one can use Maxwell's mixture equation for the
effective conductivity $\sigma_{\mathrm{eff}}$ of a suspension of
particles with volume concentration $p$ \cite{Maxwell1873}:

\begin{equation}
\label{mixeq}
\frac{\sigma_{\mathrm{eff}}-\sigma_{\mathrm{s}}}{\sigma_{\mathrm{eff}}+2\sigma_{\mathrm{s}}}=p\frac{\sigma_{\mathrm{p}}-\sigma_{\mathrm{s}}}{\sigma_{\mathrm{p}}+2\sigma_{\mathrm{s}}}
\end{equation}

\noindent Here $\sigma_{\mathrm{s}}$ and $\sigma_{\mathrm{p}}$ are
the conductivity of the solute and the particle, respectively. If we
regard the hemoglobin molecules as insulating particles (i.e.,
$\sigma_{\mathrm{p}}=0$) suspended in plasma with the same
conductivity as the extracellular plasma, we arrive at the following
ratio of solute conductivity
($\sigma_{\mathrm{s}}=\sigma_{\mathrm{a}}$) and effective
conductivity ($\sigma_{\mathrm{eff}}=\sigma_{\mathrm{i}}$)
\cite{Pauly1966}:

\begin{equation}
\label{mixeq2}
\frac{\sigma_{\mathrm{a}}}{\sigma_{\mathrm{i}}}=\frac{1+p/2}{1-p}
\end{equation}

\noindent Using $p=0.37$ (ref \citenum{Pauly1966}), a ratio of about
two is obtained, which is in quite reasonable agreement with the
findings of Fig. \ref{fig:innen}(a). The energy barrier of 0.17~eV for
charge transport within the intra- and extracellular plasma is of
the same order of magnitude as the one deduced from
$\tau_{\beta}(T)$ (0.11 - 0.15~eV, see previous section). This seems
reasonable as the temperature dependence of $\tau_{\beta}$ should be
mainly governed by the conductivity of inner and outer plasma (Eq.
\ref{taubet}). In any case one should bear in mind that the absolute
values of the energy barriers have rather high uncertainty due to
the restricted temperature range.

The dielectric constant of the cell interior
$\varepsilon_{\mathrm{i}}$ was calculated using Eq. \ref{sp2}. The
results for 310~K are listed in Table \ref{tab:Aus310} and the
temperature dependence is shown in Fig. \ref{fig:innen}(b). One
should be aware that $\varepsilon_{\mathrm{i}}$ is the dielectric
constant at frequencies below the onset of the $\gamma$-relaxation.
As expected, the obtained values are smaller than the dielectric
constant of the suspending medium
($\varepsilon_{\mathrm{a}}\approx73$ - $67$ for $T=290$ - $310$~K,
respectively), deduced from the fits of the spectra of pure plasma
(see Fig. \ref{fig:all} for 310~K). This difference is reasonable
because the main contribution to the $\varepsilon'$ of the plasma
arises from the highly dipolar water molecules
($\varepsilon_{\mathrm{s}}$ of water $\approx74$ at 310~K (ref
\citenum{Kienitz1981})) and the additional constituents of the
cytoplasma (mainly hemoglobin) should lower its permittivity. In
literature, $\varepsilon_{\mathrm{i}}$ values ranging around 40-70
were reported \cite{Pauly1966, Asami1989, Bianco1979, Asami1980}.
The calculation of $\varepsilon_{\mathrm{i}}$ by Eq. \ref{sp2}
should provide the same results for each hematocrit value. However,
as revealed by Table \ref{tab:Aus310} and Fig. \ref{fig:innen}(b),
the calculated $\varepsilon_{\mathrm{i}}$ increases by about 20\%
with increasing $Hct$, pointing out the limits of the model. Similar
behavior was also reported in ref \citenum{Bianco1979}. The decrease
of $\varepsilon_{\mathrm{i}}$ with increasing temperature revealed
by Fig. \ref{fig:innen}(b) is consistent with Curie behavior (Eq.
\ref{Curie}) expected for the dielectric strength (and thus
approximately also for the static dielectric constant) of dipolar
materials. Here one should be aware that $\varepsilon_{\mathrm{i}}$
represents the static dielectric constant of a $\gamma$-like
relaxation of the cell interior that will take place at higher
frequencies and in fact this relaxation contributes to the actually
observed $\gamma$-relaxation of blood (see section \ref{gamma}). The
uncertainty of the data in Fig. \ref{fig:innen}(b) is too large to
allow for a quantitative evaluation in terms of Eq. \ref{Curie}.
Overall, Eq. \ref{sp2} seems to lead to reasonable values of the
dielectric constant of the cytoplasma. It is based on the
determination of the plasma dielectric constant (experimentally
determined) and $\varepsilon_{\infty\beta}$, which is not influenced
by possible cell aggregation effects.

An alternative determination of the dielectric properties of the
RBCs is provided by the Hanai-Bruggemann model, Eqs. \ref{hb1} and
\ref{hb2}, which was especially proposed for highly concentrated
suspensions. From Eq. \ref{hb2} the cell conductivity was
calculated, leading, however, to negative values. Using Eq.
\ref{hb1}, the dielectric constant $\varepsilon_{\mathrm{p}}$ of the
cell is found to vary between 34 ($Hct=0.57$) and 42 ($Hct=0.39$) at
room temperature. Those values are slightly smaller than the ones
calculated from the Pauly-Schwan theory. This is a reasonable result
since the Hanai-Bruggemann model does not account for the shelled
structure of the cells and thus the obtained values represent the
average of cell membrane and interior. The dielectric constant of
the membrane can be expected to be much lower than that of the
cytoplasma (in contrast to its capacitance, which is high due to its
small thickness), which leads to the reduced values of the total
dielectric constant.

\subsection{$\gamma$-Dispersion}
\label{gamma}

\begin{figure}[h]
\centering
\includegraphics[width=8cm]{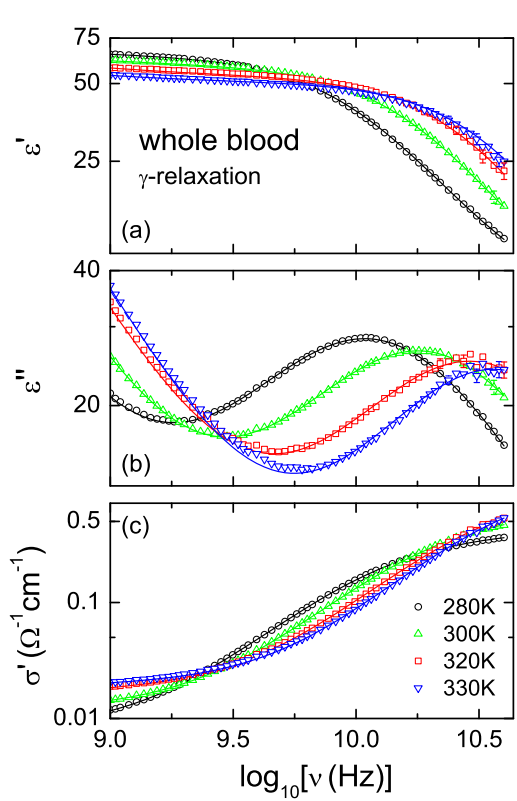}
\caption[Debye-Polarisation]{\label{gammaVollblut}
$\varepsilon'(\nu)$ (a), $\varepsilon''(\nu)$ (b), and
$\sigma'(\nu)$ (c) of whole blood ($Hct=0.39$) in the
$\gamma$-dispersion region for selected temperatures. The lines
represent fit curves as in Fig. \ref{fig:all} using the Cole-Cole
function for the description of the $\gamma$-relaxation}.
\end{figure}

The dielectric spectra of bulk water exhibit a strong relaxation
feature near 18~GHz (at room temperature) \cite{Kaatze1989}, which
is also observed in electrolytic solutions \cite{Gulich2009}. It is
commonly ascribed to the reorientational dynamics of the dipolar
water molecules (but also alternative scenarios are discussed; see,
e.g., ref \citenum{Kaatze2002}) and denoted as $\alpha$-relaxation
within the nomenclature of dipolar liquids and glass formers. The
same relaxational process also arises from the free water molecules
in blood samples. Figure \ref{gammaVollblut} shows real and
imaginary part of the permittivity (a, b) and the real part of the
conductivity (c) of whole blood in the frequency range 1 to 40~GHz
at different temperatures. The lines represent fits of the broadband
spectra as shown in Fig. \ref{fig:all}, using a Cole-Cole function
for the $\gamma$-relaxation. In $\varepsilon'(\nu)$ (Fig.
\ref{gammaVollblut}(a)), the onset of the typical relaxation steps
is seen but their high frequency plateaus,
$\varepsilon_{\infty\gamma}$, are located beyond the investigated
frequency range. Thus, exact values for
$\varepsilon_{\mathrm{\infty}\gamma}$ could not be determined and in
the fitting procedure a lower limit of 2.5 was used leading to
values between 2.5 and 6. The low-frequency plateau
$\varepsilon_{\mathrm{s}\gamma}$ of the $\gamma$-dispersion
decreases with increasing temperature. The relaxation steps and loss
peaks (Fig. \ref{gammaVollblut}(b)) show a strong
temperature-dependent frequency shift due to the slowing down of the
molecular dynamics with decreasing temperature. We find the
Cole-Cole formula (Eq. \ref{hn} with $\beta=1$) to provide the best
fits of the $\gamma$-relaxation. Figure \ref{gammaVollblut}(c) shows
the conductivity spectra with the corresponding rise and the onset
of the high frequency plateau. The low-frequency plateau of
$\sigma'(\nu)$ corresponds to the combined conductivity of plasma
and cytoplasma as discussed in section \ref{app}. The
$\gamma$-dispersion shows similar behavior for the other
investigated samples.

\begin{figure}[h]
\centering
\includegraphics[width=7.5cm]{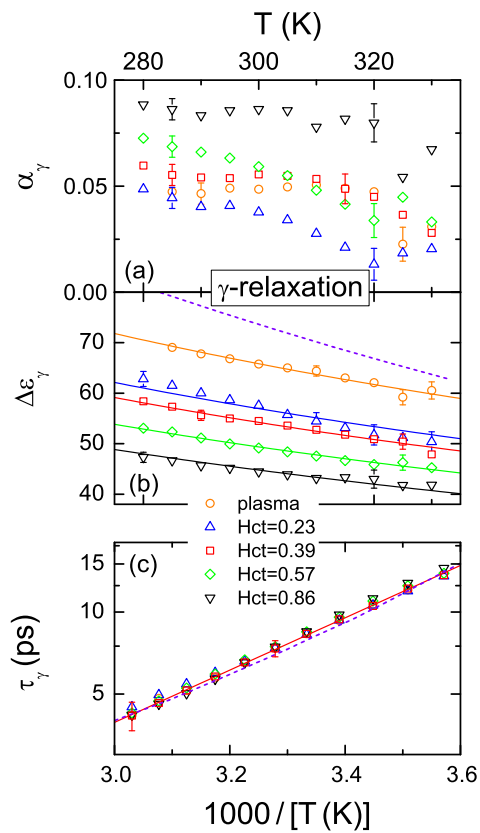}
\caption[Debye-Polarisation]{\label{gammapar}Temperature dependence
of width parameter (a), relaxation strength (b), and relaxation time
(c) as obtained from fits assuming a Cole-Cole function for the
description of the $\gamma$-relaxation (cf Fig.
\ref{gammaVollblut}). The dashed line in (b) shows literature data
for pure water (using the I.U.P.A.C. standard data for
$\varepsilon_{\mathrm{s}}(T)$) \cite{Kienitz1981}. The solid lines
in (b) are fits with a Curie-law, Eq. \ref{Curie}. The line in the
Arrhenius plot of $\tau_{\gamma}$ (c) is a linear fit of the curve
for $Hct=0.39$ (whole blood) corresponding to thermally activated
behavior, Eq. \ref{tautemp}. The dashed line shows the curve for
pure water \cite{Kaatze1989}.}
\end{figure}

The fitting parameters of the $\gamma$-relaxation,
$\alpha_{\gamma}$, $\Delta\varepsilon_{\gamma}$, and
$\tau_{\gamma}$, are provided in Fig. \ref{gammapar}. The width parameter
(a) shows a tendency to increase with increasing $Hct$. This seems
reasonable as a higher number of RBC's should lead to stronger
disorder in the system and therefore the distribution of relaxation
times should broaden. The observed decline of $\alpha_{\gamma}$ with
increasing temperature, corresponding to an approach of Debye
behavior, is a common phenomenon in dipolar liquids
\cite{Lunkenheimer2000,Schonhals1993}. It can be explained by the
growing thermal fluctuations of the environment of the water
dipoles. At very high temperatures, each dipole "sees" the time
average of the quickly fluctuating environment, which is the same
for every dipole, leading to Debye behavior \cite{Lunkenheimer2000}.

The hematocrit dependence of the relaxation strength (Fig.
\ref{gammapar}(b)) shows the expected tendency: Increasing $Hct$
values cause a decrease of the volume fraction of plasma and thus of
water in the sample, causing the reduction of the
$\gamma$-relaxation strength. The temperature dependence of
$\Delta\varepsilon_{\gamma}$ can be well parameterized by a
Curie-law, Eq. \ref{Curie}, (solid lines) with some deviations for
$Hct=0.23$ only. The obtained Curie constant, $C$, increases
smoothly from 13400 to 19800 with decreasing $Hct$. The dashed line
in Fig. \ref{gammapar}(b) corresponds to $\Delta\varepsilon(T)$ of
pure water, calculated from the I.U.P.A.C. (International Union of
Pure and Applied Chemistry) standard values for the static
permittivity $\varepsilon_{\mathrm{s}}$ of bulk water
\cite{Kienitz1981} (see also \cite{Ellison2007}) via
$\Delta\varepsilon=\varepsilon_{\mathrm{s}}-\varepsilon_{\infty}$
assuming $\varepsilon_{\infty}=4$ \cite{Hasted1973, Buchner1999}.
The relaxation strength of water matches the general trend revealed
by the other curves in Fig. \ref{gammapar}(b). However, obviously it
shows a somewhat stronger temperature dependence. This may indicate
weaker interactions between the water molecules in blood than in
pure water, which can be rationalized by the presence of the other
constituents of blood (e.g., proteins or salt ions).

Figure \ref{gammapar}(c) shows the temperature dependence of the
relaxation times $\tau_{\gamma}$ in an Arrhenius plot. The observed
linear increase is in accord with thermally activated behavior, Eq.
\ref{tautemp}. As an example, a linear fit of the data at $Hct=0.39$
is shown (solid line). From its slope an energy barrier of 0.19~eV
is obtained. There seems to be a slight increase of energy barriers
with growing $Hct$ value (from 0.18 to 0.20~eV). However, this
variation is too small to be considered significant, especially when
taking into account the rather small temperature range that could be
investigated in the present experiments due to the restricted
robustness of blood to stronger temperature variations. The present
results agree reasonably with those reported by Cook \cite{Cook1951}
who found values for $\tau_{\gamma}$ of whole blood varying between
11.9 and 7.0~ps at three temperatures between 298 and 308~K. Gabriel
et al.\cite{Gabriel1996b} reported 8.4~ps at 310~K, about $30\%$
higher than our result of 6.5~ps. The dashed line in Fig.
\ref{gammapar}(c) represents $\tau(T)$ of pure water as measured by
Kaatze \cite{Kaatze1989}. Obviously the $\gamma$-relaxations in the
investigated blood samples exhibit nearly identical dynamics as the
main relaxation of pure water. Water shows some small deviations
from Arrhenius behavior, which seem to be absent in blood but these
differences are of limited significance. However, one may speculate
that the non-Arrhenius behavior of water arises from increasing
cooperativity of the molecular motions at low temperatures as often
invoked to explain corresponding findings in glass forming liquids
\cite{Ediger1996, Ngai2000}. In blood, its other constituents can be
expected to lead to a reduction of the direct interactions between
the water molecules and thus to less cooperativity. In addition,
there are speculations of a first-order phase transition in
supercooled water \cite{Angell2008}, which may lead to critical
power-law behavior even in the normal liquid state, thus also
explaining the deviation from Arrhenius behavior in Fig.
\ref{gammapar}(c). Both scenarios are consistent with the different
temperature dependence of $\Delta\varepsilon(T)$ of water and blood
discussed in the previous paragraph.

\subsection{Further Dispersions}
\label{further}

\begin{figure}[h]
\centering
\includegraphics[width=8cm]{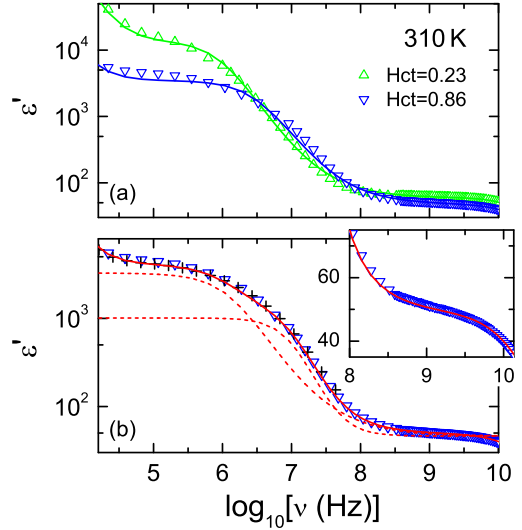}
\caption[Debye-Polarisation]{\label{deltaneu}(a) Comparison of
$\varepsilon'(\nu)$ of the blood samples with the lowest
($Hct=0.23$) and highest hematocrit value ($Hct=0.86$) in the
frequency range of the $\beta$- and $\delta$-dispersion. The lines
represent fit curves as in Fig. \ref{fig:all} and Fig.
\ref{betaVollblut} using the Cole-Cole function for the description
of the $\beta$-relaxation. (b) $\varepsilon'(\nu)$ for $Hct=0.86$
(triangles: same data as in (a), crosses: measurement with different
experimental setup). The solid line in (b) shows an alternative fit
with two Cole-Cole functions for the $\beta$-relaxation. The two
separate relaxation steps are indicated by the dashed lines. The
inset shows a magnified view of the high-frequency region.}
\end{figure}

In section \ref{beta} it was shown that fits using the Cole-Cole
function provide a reasonable description of the $\beta$-relaxation
region. However, there are some minor deviations of fits and
experimental data especially for the higher hematocrit values. This
is demonstrated in Fig. \ref{deltaneu}(a) where dielectric-constant
data for the blood samples with the highest and lowest hematocrit
values are shown. In contrast to the 23\% sample, the fit of the
spectrum of the highly concentrated sample clearly is of inferior
quality. Similar deviations were previously also observed in
Cole-Cole fits of data on disc-shaped rabbit-erythrocyte suspensions
\cite{Hayashi2008}. A close inspection of the spectrum at $Hct=0.86$
(Fig. \ref{deltaneu}(b)) seems to indicate that it may be composed
of two separate relaxation steps. However, one could suspect an
experimental artifact because in the $\beta$-dispersion region the
spectrum is composed of results from two different experimental
methods with the transition close to 10~MHz (see section
\ref{Materials}). To exclude this, in Fig. \ref{deltaneu}(b)
additional data extending from 100~kHz to 50~MHz obtained with a
different apparatus (autobalance bridge Agilent 4294A) are shown,
which exactly reproduce the two other data sets. Thus, a fit using
the sum of two separate relaxation contributions was performed
(solid line in Fig. \ref{deltaneu}(b)). It provides an excellent
description of the spectrum revealing relaxation times of $16$~ns
and $146$~ns. A very similar fit using two Cole-Cole functions was
shown by Asami and Yamaguchi \cite{Asami1999} to provide a good
description of data on human erythrocyte suspensions.

In blood there are various possibilities for additional relaxational
processes, in addition to those considered for the explanation of
the $\alpha$-, $\beta$-, and $\gamma$-relaxation: (i) the
reorientation of protein-bound water molecules, (ii) the hemoglobin
$\beta$-relaxation (i.e., the tumbling motion of the protein
molecules), (iii) the motion of polar protein subgroups, (iv) the
Maxwell-Wagner relaxation of the cell interior, or (v) the
additional Maxwell-Wagner relaxations due to the non-spherical cell
shape, to name just the most likely ones. Most of them can be simply
excluded based on the very large amplitude of
$\Delta\varepsilon_{\mathrm{s}}\approx1000$ of the additional
relaxation suggested by the fit shown in Fig. \ref{deltaneu}(b): (i)
Bound water cannot have a larger $\Delta\varepsilon$ than free
water. (ii) The hemoglobin $\beta$-relaxation in aqueous solution
was found to have a $\Delta\varepsilon$ of the order of 100
\cite{Schwan1957a, Schwan1983, Pennock1969}. It seems unreasonable
that it should be higher in the hemoglobin/cytoplasma solution of
the cell interior. (iii) Polar protein subgroups can be expected to
have smaller relaxation strength that the main tumbling relaxation.
(iv) In principle, the capacitance of the cell interior should also
be shorted at high frequencies and, indeed, the Maxwell-Wagner model
of shelled particles predicts a corresponding relaxation
\cite{Asami2002}. However, for any reasonable choice of parameters
this capacitance is too small to lead to any considerable
contribution to $\varepsilon'$ and this additional relaxation
usually is considered negligible \cite{Schwan1957a, Schwan1983, Pennock1969}.
Thus, the non-spherical shape of the RBCs seems to be
the most likely cause of the additional relaxation observed in
Fig. \ref{deltaneu}(b). Already in ref \citenum{Schwan1957a} deviations
from simple relaxation behavior of erythrocyte suspensions were
ascribed to the non-spherical form of RBCs and also Asami and
Yamaguchi \cite{Asami1999} explained their results in this way. As
mentioned in section \ref{cellmod}, the Maxwell-Wagner model for
suspensions of spheroid particles predicts up to four relaxations
\cite{Asami2002} (two of them arise from the cell interior and can
be neglected). However, the found relaxation-time ratio of the order
of 10 is too high to be explainable by this model, at least if
assuming a reasonable ratio of the two semi-diameters of the
spheroids \cite{Asami2002, Asami2002a}. The spectra on rabbit
erythrocytes, mentioned above, also could not be described by the
Maxwell-Wagner model for spheroid particles \cite{Hayashi2008}.
However, one should be aware that RBCs only roughly can be
approximated by spheroids and most likely their biconcave shape
plays a role in the found discrepancies.

An additional $\delta$-dispersion between $\beta$- and
$\gamma$-relaxation is often invoked to explain a slow continuous
decrease of $\varepsilon'(T)$, observed in the region from several
10~MHz to about 3~GHz in various biological materials, including
protein solutions \cite{Schwan1957a, Schwan1965, Pennock1969,
Grant1966, Grant1974} and blood \cite{Schwan1957a}. It has been
ascribed to various mechanisms as the dynamics of protein-bound
water molecules or polar subgroups of proteins. Indeed such a
dispersion is also found in our present results on blood and becomes
most pronounced for the high hematocrit values (see inset of Fig.
\ref{deltaneu} for an example). However, it can be completely
described by the broadband fits promoted in the previous sections
(line in inset), especially if including a second relaxation in the
$\beta$-relaxation region as shown in Fig. \ref{deltaneu}(b). Thus
in our blood samples the apparent dispersion in this region arises
from the superposition of $\beta$- and $\gamma$-relaxation and we
find no evidence for a $\delta$-relaxation. However, the presence of
a small $\delta$-relaxation is not completely excluded by this
finding.

\section{Summary and Conclusions}

In the present work, we have provided dielectric spectra of human
blood for an exceptionally broad frequency range and at different
temperatures and hematocrit values. A combination of models for the
different dispersion regions enabled nearly perfect fits of the
broadband spectra. The obtained fit curves represent an excellent
estimate of the dielectric properties of blood for a wide range of
parameters. They are provided for electronic download in the
supporting information and can be employed for SAR calculations and
other application purposes. The different dispersion regions have
been analyzed in detail. The observed electrode-polarization effects
are accounted for by an equivalent circuit model assuming a
distribution of relaxation times. While our analysis of the
low-frequency region cannot completely rule out the presence of an
$\alpha$-dispersion in blood, we can satisfactorily describe our
low-frequency data without invoking such a relaxation. This finding
agrees with earlier results stating the absence of an
$\alpha$-relaxation in blood \cite{Bothwell1956}. The analysis of
the $\beta$-relaxation using standard cell models partly leads to
unreasonable results for the intrinsic dielectric properties. This
most likely can be ascribed to cell aggregation playing an important
role in "real" blood samples, in contrast to suspensions of
erythrocytes prepared by standard solutes, often reported in
literature. Cell aggregation seems to be important especially for
the dielectric behavior at the low-frequency side of the
$\beta$-relaxation. In contrast, using only parameters determined at
frequencies beyond the $\beta$-peak frequency, leads to reasonable
estimates of the conductivity and dielectric constant of the cell
interior. In addition, we find strong hints that the
$\beta$-relaxation is in fact composed of two separate relaxation
processes, which we ascribe to the marked deviations of the RBCs
from spherical geometry.

We observe clear dispersion effects in the region between the
$\beta$- and $\gamma$-relaxation, which often is ascribed to a
so-called $\delta$-relaxation. However, our analysis of the
broadband spectra including electrode polarization,
$\beta$-dispersion, and $\gamma$-relaxation leads to excellent fits
in this region, which thus is revealed to be a superposition of
different contributions and not due to a separate relaxation
process. Thus, while there clearly is dispersion in blood between
the $\beta$- and $\gamma$-relaxation, there is no evidence for a
$\delta$-relaxation. Finally, detailed information on the
$\gamma$-relaxation in blood is provided. Its properties closely
resemble those of the relaxation caused by reorientational molecular
motions in pure water. However, some minor differences arise, which
seem to indicate less cooperative motions of the water molecules in
blood samples.

Overall, the dielectric spectra of blood are of astonishing
simplicity if considering the complexity of blood, being composed of
a variety of different constituents. In fact we have described our
broadband spectra without assuming any intrinsic frequency
dependence in the complete range from 1~Hz up to about 1~GHz and
only the $\gamma$-dispersion arising from the tumbling motion of the
water molecules is of intrinsic nature. Of course, there is the
strong $\beta$-relaxation, which may be regarded as quasi-intrinsic
but as it is of Maxwell-Wagner type, in a narrower sense it should
be considered as artificial. However, of course for many
applications (e.g., the calculation of SAR values) the overall
dielectric properties and not only the intrinsic ones are of
essential importance and we hope our work will serve for these
purposes in the future.

\section{Acknowledgements} We gratefully acknowledge the help of Dr.
W. Behr, Dr. K. Doukas, and Prof. Dr. W. Ehret at the "Klinikum
Augsburg" in the taking and preparing of the blood samples.

\appendix
\section{Supplementary data}

\noindent Fit curves of the broadband spectra and fit parameters for
all samples and temperatures investigated in the present work are
available for electronic download.







\bibliographystyle{model1-num-names}








\end{document}